\documentclass[twocolumn,english,showpacs]{revtex4}
\usepackage[T1]{fontenc}
\usepackage[latin9]{inputenc}
\usepackage{graphicx}
\usepackage{amssymb}

\usepackage{mathrsfs}

\usepackage{float}

\usepackage{indentfirst}

\makeatletter

\def\journal #1, #2, #3, 1#4#5#6{{\sl #1~}{\bf #2}, #3 (1#4#5#6) }

\usepackage{mathrsfs}
\usepackage{float}
\usepackage{indentfirst}

\def\eqa{\begin{eqnarray}}
\def\eea{\end{eqnarray}}
\newcommand{\eq}{ \begin{equation}}
\newcommand{\ee}{\end{equation}}
\newcommand{\nn}{\nonumber\\}

\renewcommand{\Im}{{\rm Im}}

\newcommand{\ua}{\uparrow}
\newcommand{\da}{\downarrow}

\makeatother

\usepackage{babel}

\begin{document}

\title{Spin current through an ESR quantum dot: A real time study}

\author{Lei Wang$^{1}$, Hua Jiang$^{1}$, J. N. Zhuang$^{1}$, Xi Dai$^{1}$, and X. C. Xie$^{2,1}$}

\address{$^{1}$Beijing National Lab for Condensed Matter Physics and Institute of Physics,
Chinese Academy of Sciences, Beijing 100190, China }
\address{$^{2}$Department of Physics, Oklahoma State University, Stillwater, Oklahoma 74078, USA}

\begin{abstract}
The spin transport in a strongly interacting spin-pump nano-device
is studied using the time-dependent variational-matrix-product-state
(VMPS) approach. The precession magnetic field generates a
dissipationless spin current through the quantum dot. We compute the
real time spin current away from the equilibrium condition. Both
transient and stationary states are reached in the simulation. The
essentially exact results are compared with those from the
Hartree-Fock approximation (HFA). It is found that correlation
effect on the physical quantities at quasi-steady state are captured
well by the HFA for small interaction strength. However the HFA
misses many features in the real time dynamics. Results reported
here may shed light on the understanding of the ultra-fast processes
as well as the interplay of the non-equilibrium and strongly
correlated effect in the transport properties.
\end{abstract}




\maketitle

\section{Introduction}

The rapid progress of nano-electronics and information technologies
has prompted intense interest in exploiting the spin properties of
the electrons, which results in the emergence of
spintronics.\cite{Wolf:2001p7356} One of the most important
spin-based electronic devices is a mesoscopic quantum dot (QD)
system. Spin-polarized transport through a QD has been extensively
investigated recently. It has been shown theoretically
\cite{Mucciolo:2001p7357} and demonstrated experimentally
\cite{Watson:2003p7358} that a QD system will function as a
phase-coherent spin pump in the presence of sizable Zeeman
splitting. Very recently, spin-polarized current has been detected
from a quantum point contact (QPC) \cite{Potok:2002p7066} and from
Coulomb-blocked QDs \cite{Potok:2003p7065}. The electron spin
resonance QD (ESR-QD) could also serve as an element device for
quantum computing \cite{Koppens:2006p7200}.

Due to its small size, Coulomb correlation could play important role
in the transport experiments involving a QD. At low temperature,
Kondo effect creates new states with many-body character at the
Fermi level. Although the effect of Kondo resonance on the charge
current has been studied in the QD for different situations, its
influence on the spin current is less known. The physical processes
involved in the transport experiments are out of equilibrium in many
cases. Moreover, the state of the system may also be time dependent.
These features make the investigations unaccessible from the
conventional many-body tools. So an essentially exact numerical
method for the non-equilibrium and time dependent phenomena in the
interacting nano-devices is highly desirable, which can also verify
the approximations used in various analytical approaches.

There exist several powerful methods to deal with the low
dimensional correlated systems, such as the numerical
renormalization group (NRG) and the density-matrix renormalization
group (DMRG). With the input from the quantum information science,
time-evolving block decimation (TEBD)
\cite{Vidal:2004p5978,Vidal:2003p5980} and adaptive time dependent
DMRG (t-DMRG) have received much attention.\cite{White:2004p5977,
Daley:2004p5979} To our knowledge, there are some previous studies
of the non-equilibrium transport of the nano-devices using the
adaptive t-DMRG technique.\cite{AlHassanieh:2006p6942, Kirino:2008p6951,
HeidrichMeisner:2009p12615} By adopting the logarithmic
discretization, \citet{Silva:2008p7061} have studied the Kondo
correlations. \citet{Guo:2009p7010} have examined the noninteracting
resonant level model in the Landau-Zener potentials. Besides the
adaptive tDMRG approach, there are also attempts based on the
time-dependent NRG \cite{Anders:2005p6185}, functional RG
\cite{Karrasch:2006p4860}, Dyson equation embedding
\cite{HeidrichMeisner:2009p7068}, flow-equations
\cite{Kehrein:2005p8744,Lobaskin:2005p8743}, and quantum Monte
Carlo\cite{Schiro:2009p7094, Schmidt:2008p6350,Werner:2009p7093}
methods. But the adaptive t-DMRG approach gives direct access to the
transient regime and could handle the time dependent Hamiltonian
directly.  The tVMPS approach adopted in this paper is directly
connected with the adaptive t-DMRG \cite{White:2004p5977,
Daley:2004p5979} and the TEBD \cite{Vidal:2004p5978,Vidal:2003p5980}
approaches. The computational cost of these two methods is very
similar, and in practice both methods achieve a similar accuracy.
\cite{Verstraete:2008p5609}.

The merit of the VMPS approach lies in two aspects. First, it
represents a large class of states, which could be seen from the
success of the NRG and DMRG approaches (which generate MPS in their
processes) for the zero and one-dimensional quantum models. And the
fact that the entanglement entropy increases slowly in one dimension
also permits one to simulate the states classically using the VMPS
method. \cite{Verstraete:2008p5609}. Second, it is easier to handle
the MPS \textit{i.e.}, the overlap of two MPS, the expected value of
an operator in a given MPS \textit{etc.} can be calculated with
polynomial complexity.

In this paper we study the spin current through an ESR quantum dot
with Coulomb interaction
\cite{Zhang:2003p1086,Hattori:2007p12613,Hattori:2008p11172}. We
obtain the transient as well as the quasi-steady state spin current
using the time-dependent VMPS method. We also studied the effect of
the interacting on the spin current. The results are compared with
time-dependent Hatree-Fock approximation (TD-HF) and non-equilibrium
Green's function (NEGF) approach for the quasi-steady state.

\section{Model}

We consider an ESR setup of a quantum dot, where single-electron level of the dot
is split by the Zeeman field $B_0$ and the two spin levels are coupled by a
rotating magnetic field $B_1(\cos(\omega \tau), \sin(\omega \tau))$.
The Hamiltonian reads $H=H_{\mathrm{dot}}+ H_{\mathrm{rotate}} + H_{\mathrm{lead}} $

\begin{eqnarray}
H_{{\mathrm{dot}}} &=& V_{g} \sum_{\sigma} n_{\sigma} -
\frac{g\mu_{B}B_{0}}{2} (n_{\ua}-n_{\da}) +  U n_{\ua} n_{\da}
\nonumber\\H_{\mathrm{rotate}} &= &  -\frac{g\mu_{B}B_{1}}{2}
(d_{\ua}^{\dagger}d_{\da}e^{i\omega \tau} +
d_{\da}^{\dagger}d_{\ua}e^{-i\omega \tau})                  \nn
H_{\mathrm{lead}} &=& - t^{\prime} \sum_{\sigma}
(d_{\sigma}^{\dagger} c_{1,\sigma}+h.c.)\nonumber \\
&& -t\sum_
{i=1;\sigma}^{N_{lead}} (c_{i,\sigma}^{\dagger} c_{i+1,\sigma} ~+~
h.c. )\,.\label{eq:ham}
\end{eqnarray}

Here $d$ and $c$ denote the annihilation operator of electron in the
dot and the lead. $H_{\mathrm{lead}}$ contains the terms describing
coupling of the dot with the lead and
 the hopping in the lead. $H_{\mathrm{rotate}}$ contains the rotating magnetic field.
 $H_{\mathrm{dot}}$ contains the Zeeman splitting, the gate voltage terms and
 the on site Coulomb repulsion between the spin up and down elections.
 In the present study we fix $t=1$ and $g\mu_{B}B_{0}=\omega=1$
 where the ESR resonance condition is satisfied.
 We set $g\mu_{B}B_{1}=2$ and $t^{\prime}$=0.4 unless mentioned.
 The single lead is a noninteracting chain with $N_{lead}$ sites.
 The coupling of it with the QD is described by the hybridization $\Gamma=\pi {t^{\prime }}^2 \rho$.

\begin{figure}
    [tbp] \centering
    \includegraphics[height=4cm, width=8cm]{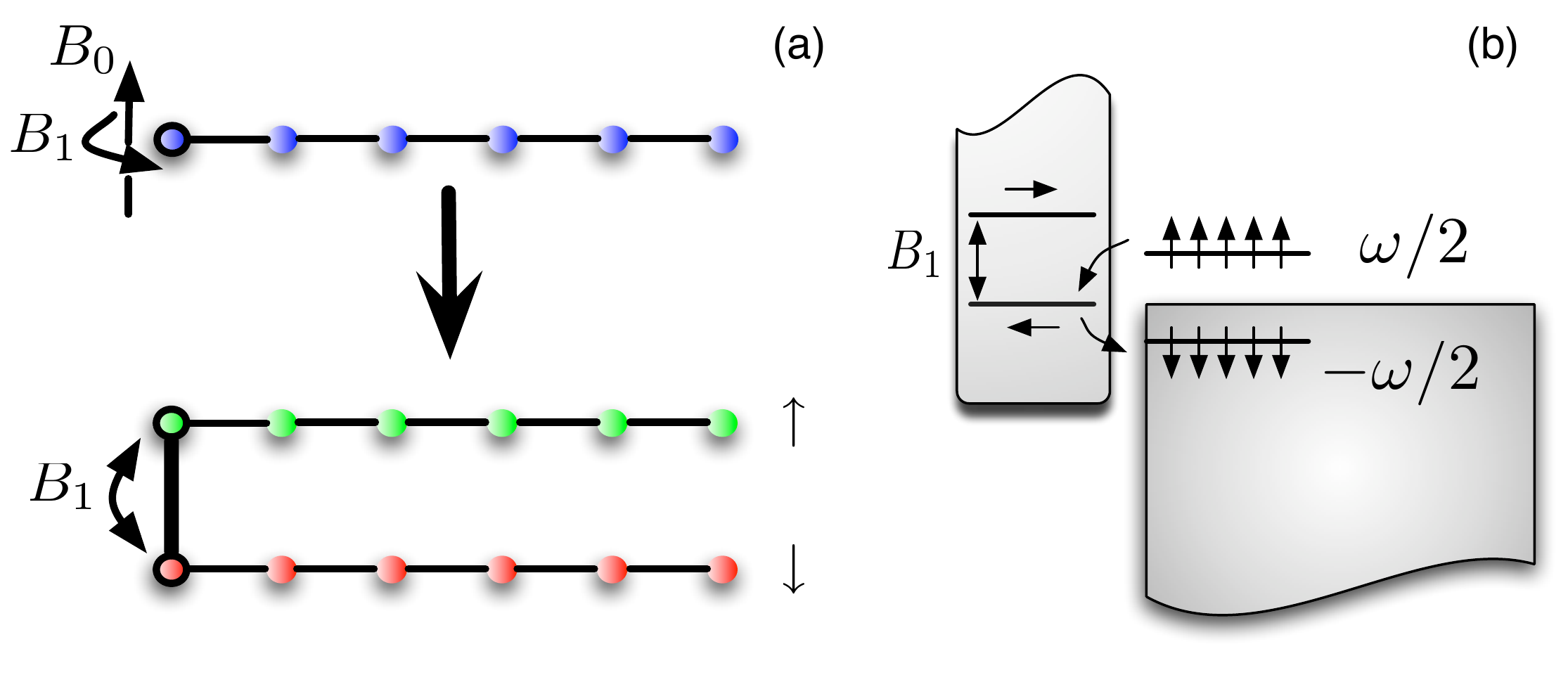}
    \caption{(a).
    The experimental set up of an ESR quantum dot. Besides the Zeeman field $B_{0}$
    along the $z$-axis, there is a rotational magnetic field $B_1$ acting in the $xy$-plane.
    The dot is coupled to a noninteracting lead and forms a single impurity Anderson model (SIAM) chain.
    The SIAM chain is unfolded into two spinless chains, and they are coupled at the leftmost end
    by Coulomb interaction $U$ and the rotational magnetic field $B_1$. (b). Schematic view of
    time independent energy levels in the rotational reference frame. The Fermi level of the
    spin up/down electrons in the lead is shifted to $\pm\omega/2$, while the $|\leftarrow\rangle$
    and $|\rightarrow\rangle$ levels in the dot split by $2g\mu_{B}B_{1}$. The dot levels could be
    tuned by the gate voltage $V_{g}$. } \label{fig:unfolded_rfm}
\end{figure}

\section{Methods}

The time dependence of the Hamilton could be eliminated by the
unitary transformation $\mathcal{U} = e ^ {- i (\omega
\tau/2)[\sum_{i}(c_{i,\da}^{\dagger}c_{i,\da} +
c_{i,\ua}^{\dagger}c_{i,\ua}) + (d_{\da}^{\dagger}d_{\da} -
d_{\ua}^{\dagger}d_{\ua}) ]}$. It transforms the Hamiltonian into
rotating reference frame (RF), see Fig \ref{fig:unfolded_rfm}(b).
One can see that the rotating magnetic field, in effect, shifts lead
electron energy to the opposite directions for up and down spins.
Quasi-steady state spin current has been studied in the RF using the
non-equilibrium Green's function (NEGF) method in the noninteracting and
infinite $U$ case \cite{Zhang:2003p1086} and by NRG approach \cite{Hattori:2008p11172} in adiabatic limit. In the following we will
revisit the problem in transient as well as quasi-steady regime, and treat
the interacting non-perturbatively with the exact numerical methods.

\begin{figure}
    [tbp] \centering
   \includegraphics[height=6cm, width=8cm]{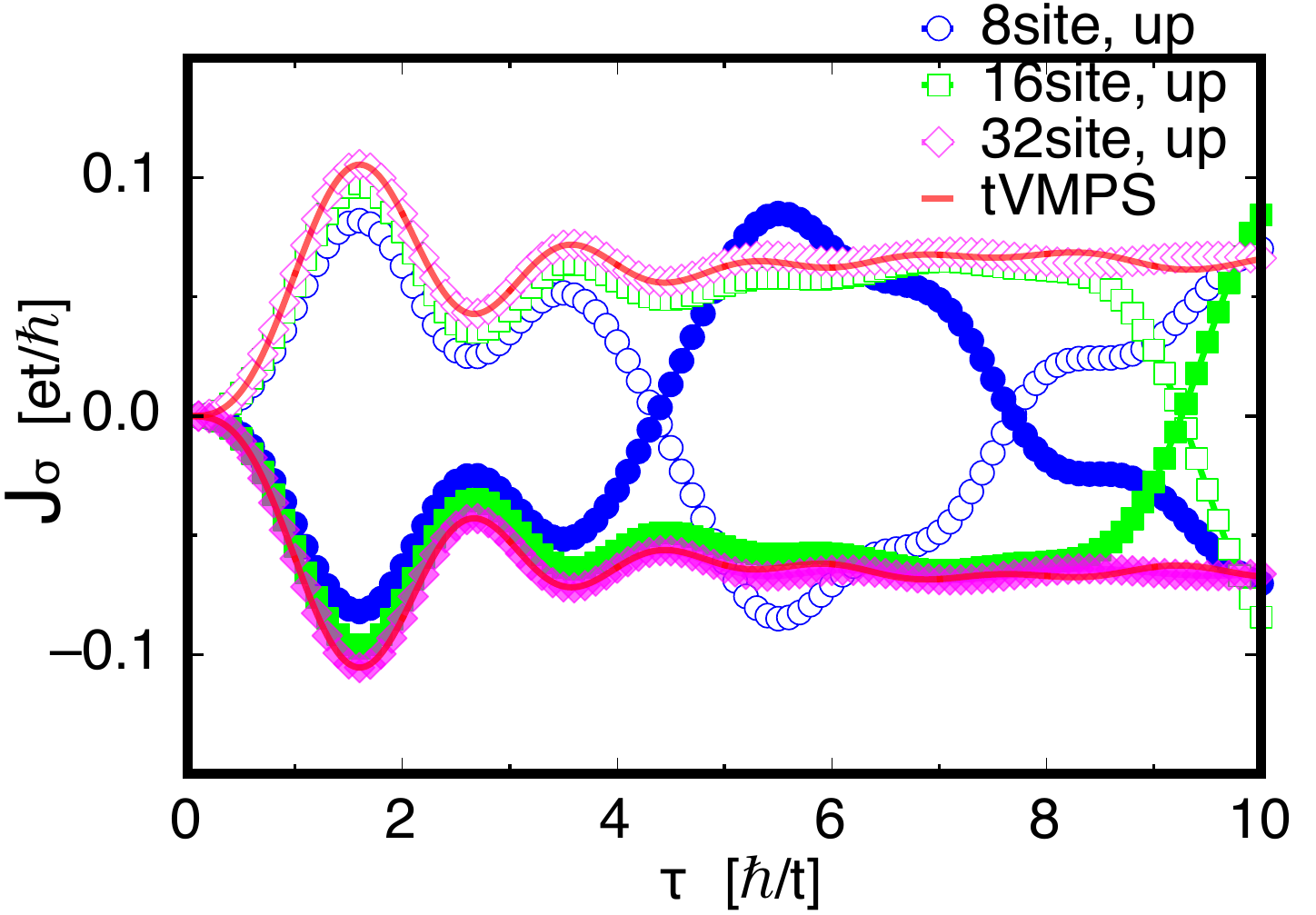}
  \caption{(Color online)
    The spin current for different chain length $L=8,16,32$. The red solid line
    is the tVMPS result and the dots are from direct integration of the von-Neumann equation,
    see text. Quasi-steady state could be recognized for larger chain length.
    Where $J_{\ua}$ (hollow dots) and $J_{\da}$ (filled dots) are of opposite sign. } \label{fig:nonint}
\end{figure}

For the time dependent Hamiltonian $H(\tau)$, the evolution of the
density matrix $\rho$ follows the von-Neumann equation $i \frac{d \rho}{d
\tau} = [H, \rho]$. For the noninteracting case, one could write
$H(\tau)$ with the single particle basis. Then direct integration of
the von-Neumann equation with a given initial condition
$\rho(\tau)\vert_{\tau=0}=\rho_{0}$ could be easily done. The
initial density matrix $\rho_{0}$ is calculated form ground state of
the SIAM chain without the rotating magnetic field $B_{1}$.

From the density operator $\rho(\tau)$ the occupation on the dot is calculated
by $n_{\sigma}(\tau) = \langle \psi(\tau)|d^{\dagger}_{\sigma}d_{\sigma}|\psi(\tau) \rangle$,
and the spin current through the dot is $J^{\sigma} = \frac{2 e}{\hbar} \Im [t^{\prime} \langle
\psi(\tau)|d_{\sigma}^\dagger c_{1,\sigma} |\psi(\tau) \rangle]$, \textit{i.e.}
they are evaluated on the bond connect the dot and the lead. We set $g\mu_{B}=e=\hbar=1$ in this paper.

For the interacting case we adopt the VMPS approach to calculate the
ground state of $H(\tau=0)$. Then we apply the rotational magnetic
filed $B_{1}$ on the dot at $\tau=0$, and perform the time
evolution.\cite{Verstraete:2004p5961}. To reduce the dimension of
the local Hilbert space, an unfolded technique
\cite{Saberi:2008p5625} is used, the original SIAM chain is unfolded
into two chains with different spins. The total length of the
unfolded SIAM chain is $L$. They are connected at the end point by
the Coulomb repulsion and the rotating magnetic field $B_{1}$. The
errors of the computation mainly come from the following sources.
First the Trotter decomposition error. Second the truncation errors
accumulated in the course of the time evolution. For short time
scale the Trotter error dominates while for the long time the
truncation error dominates. Caution must also be taken because of
the finite size of the leads. The electrons may bounce at the end
point and the spin current flows along the reverse direction. This
is an artifact of the present method and could be eliminated by
careful finite size scaling analysis.

We also use time-dependent Hartree-Fock (TD-HF) method to investigate the interacting case approximately,
in which the interaction term is factorized into $U (\langle n_{\uparrow}\rangle n_{\downarrow}
+ n_{\uparrow}\langle n_{\downarrow}\rangle )$. At each time step of the integration of the von-Neumann equation, the dot occupation which is used to update the Hartree-Fock Hamiltonian.
This approach is as efficient as the non-interacting cases.

\section{Results}

\begin{figure}
    [tbp] \centering
   \includegraphics[height=7cm, width=8cm]{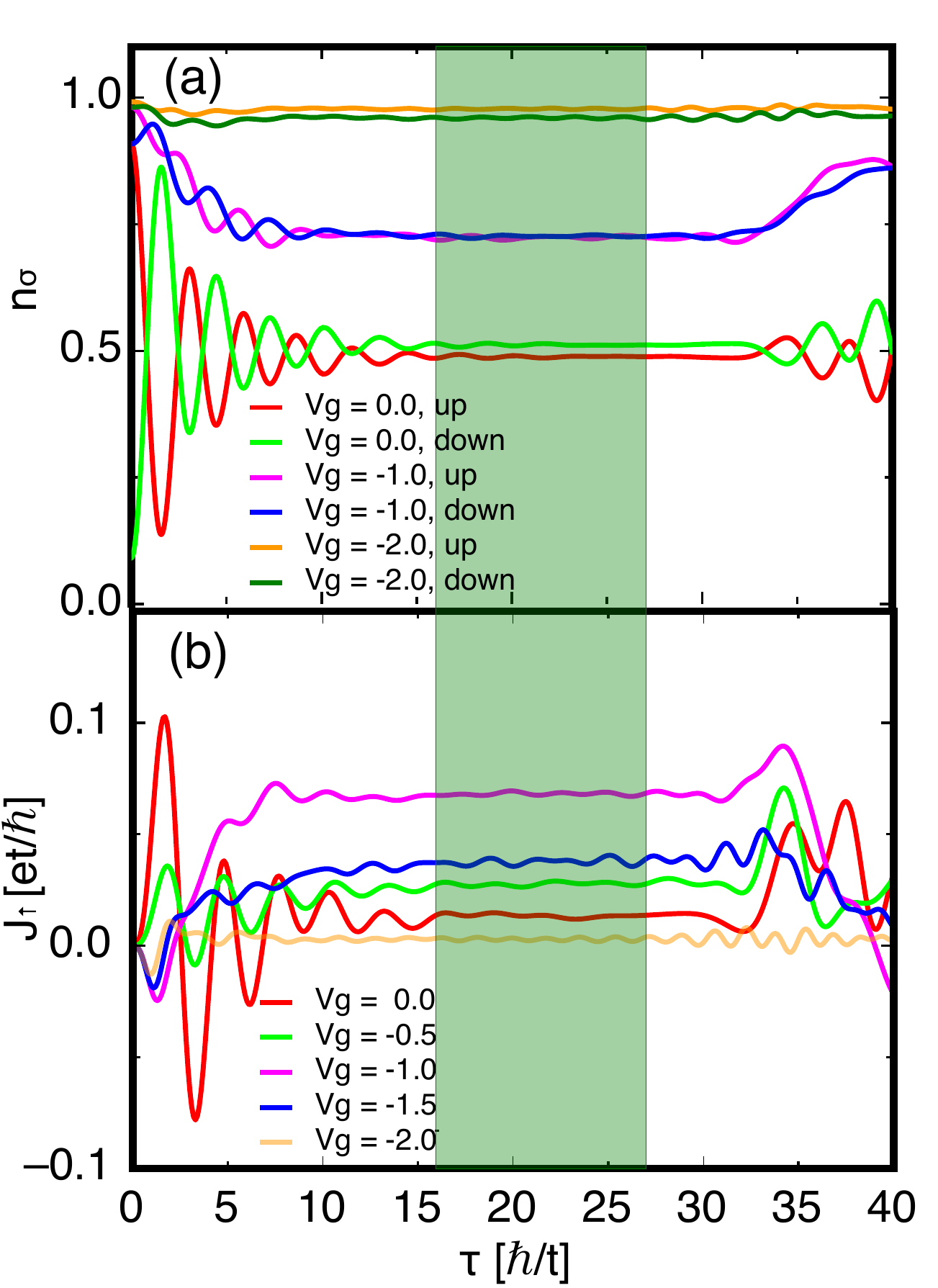}
   \caption{(color online)
    Dot occupancy for each spin (upper panel) and spin current (lower panel)
    for a $64$ sites chain with different gate voltage. The shaded region is
    averaged to calculate the quasi-steady state spin current.} \label{fig:Javerage}
\end{figure}

In Fig. \ref{fig:nonint} we show the development of the spin current
after the rotational magnetic field $B_{1}$ is applied to the
noninteracting dot. To reduce the transient region, we use a larger
coupling constant $t^{\prime}=1$ here. For small chain length
($L=8,16$) the finite size effect shows up within the maximum time
of our simulation. The charge pulse reaches the end of the chain and
bounces back thus causes the reversing of the sign of the spin
current. However, the spin current reaches a saturation value for
larger chain length ($L=8,16$) which indicates the quasi-steady
state spin transport is achieved. In the quasi-steady state, the
spin up electrons flow in the dot, flip their spins and then flow
away. There is no net charge transport since $J_{\ua}=-J_{\da}$.
Overshooting behavior at short time scales ($\tau \sim 1.8$) is
observed. It is due to abrupt applying of the rotational magnetic
field and could be suppressed if the rotational magnetic field is
turned on adiabatically. The tVMPS result for $L=32$ nicely follows
the direct integration of the von-Neumann equation. The oscillation
at long times in the tVMPS approach was also noticed in the previous
study with the adaptive t-DMRG approach \cite{AlHassanieh:2006p6942}.
With the increasing of the bond dimension, the oscillation tends to
disappear. The coincidence validates the VMPS method for the
noninteracting case. However, its main power lies in the interacting
cases and we expect similar precision could be achieved in that
case.

\begin{figure}[tbp] \centering
   \includegraphics[height=7cm, width=8cm]{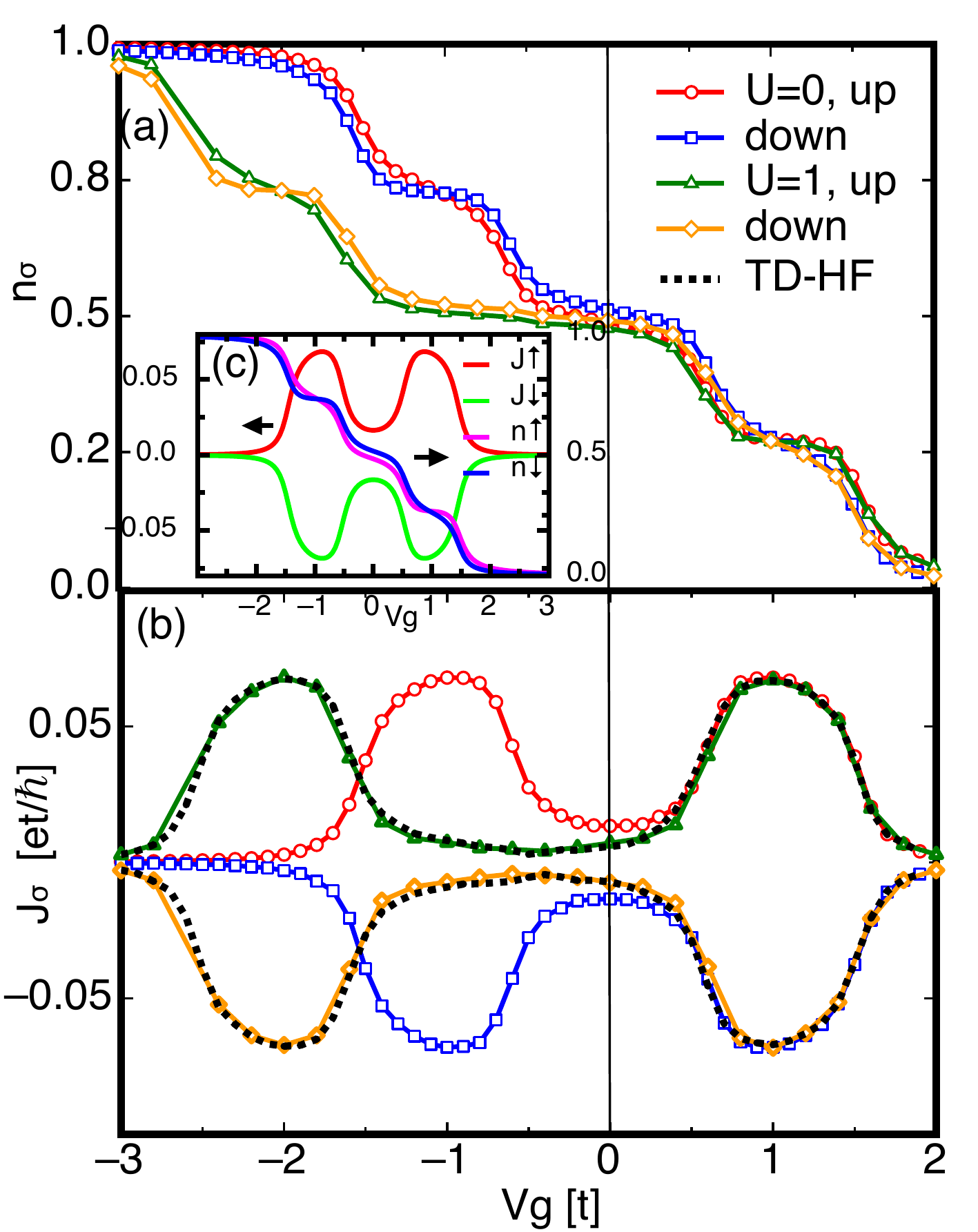}
  \caption{(Color online)
    Occupation of electrons on the dot (upper panel) and spin current through the dot (lower panel)
    for different gate voltages. The spin current of high filling region is not affected by the interacting,
    while the spin current peak at low filling is shifted by $U$.
    Inset are the $U=0$ quantities calculated in rotational reference frame by NEGF approach.} \label{fig:varyVg}
\end{figure}

Electron occupation and spin current through the quantum dot are
calculated as functions of time for different gate voltages $V_{g}$,
shown in Fig.\ref{fig:Javerage}. It is seen that the overall
occupation on the dot oscillates as it reaches its
quasi-steady-state value, the spin current develops in the mean
time. Note that the transient state current can even be of the
opposite sign with its value in the quasi-steady state,
Fig.\ref{fig:Javerage}(b). By taking average of physical quantities
in the quasi-steady state, we extract variation of the
quasi-steady-state dot occupation and spin current with the gate
voltage $V_{g}$, see Fig.\ref{fig:varyVg}. The spin polarization in
the steady state is less pronounced than the initial state. And it
is even inverted for $-1<V_{g}<1$, Fig.\ref{fig:varyVg}(a). The
steady state current attains its maximum value at $V_g=\pm 1$. This
is the case where $n_{\ua}=n_{\da}$ \textit{i.e.} the dot is spin
unpolarized. It can be seen from Fig.\ref{fig:unfolded_rfm}(b), in
this case $V_{g}\pm B_{1}=0$ and the $|\leftarrow\rangle$
($|\rightarrow\rangle$) is tuned to zero energy point. Spin flip
process on the dot is most efficient and gives a maximum spin
current. We also use NEGF to study the current and occupation in the
RF, see inset of Fig.\ref{fig:varyVg}(a). It gives qualitatively
similar results, the small discrepancy is due to the finite size
effect of the tVMPS approach. It should be noted that NEGF approach
adopted here does not yield results of transient state since the
unitary transformation eliminates the time dependence. To take into
account the initial condition properly, the two-time Green's
function should be used \cite{Myohanen:2008p12607,
Balzer:2009p12608,Myohanen:2009p12606}.

\begin{figure}
    [tbp] \centering
   \includegraphics[height=7cm, width=8cm]{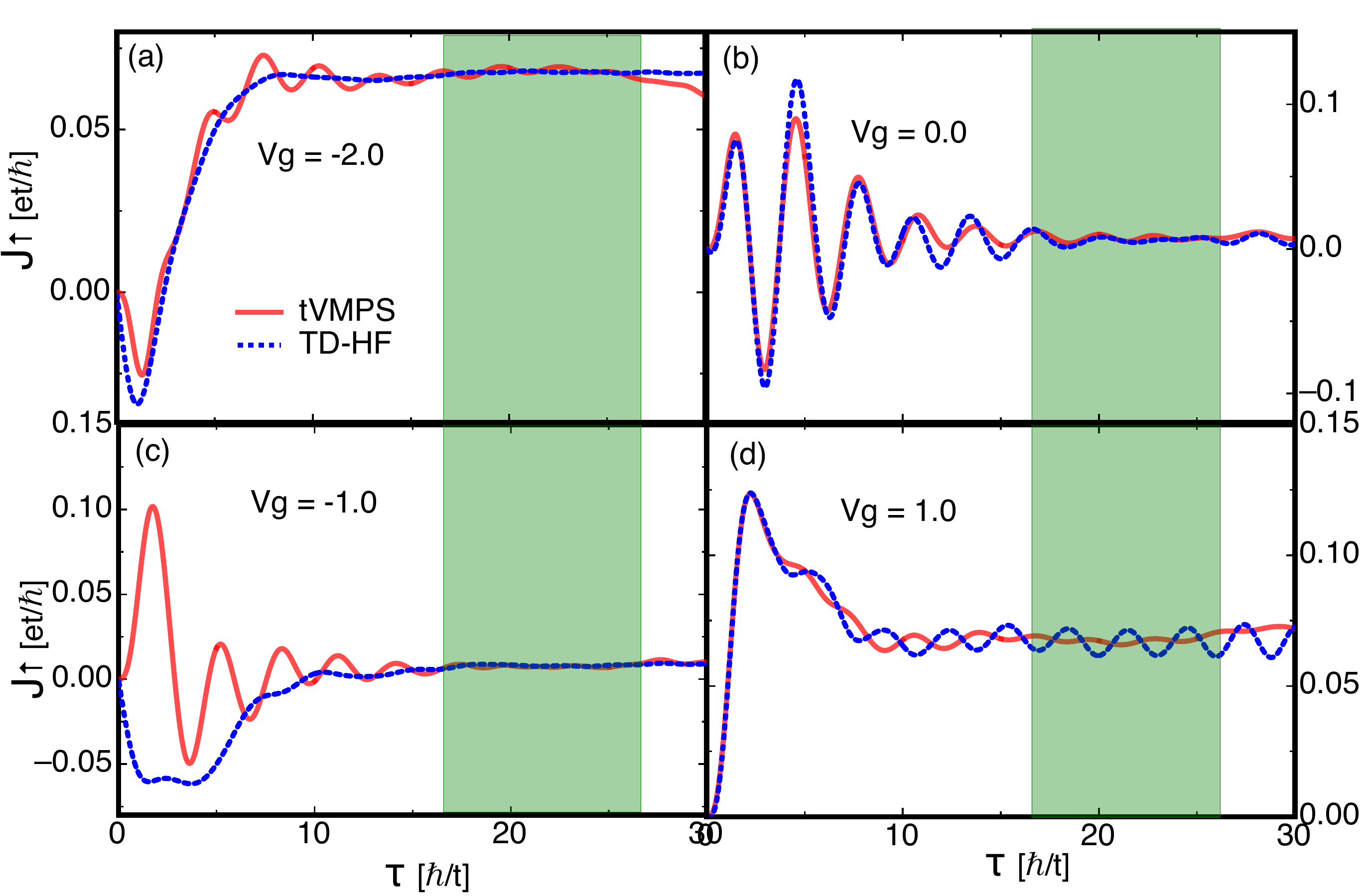}
   \caption{(color online)
    The comparison of the tVMPS (full line) and TD-HF (dot line) results of spin current for different gate voltage, where the interacting strength $U=1$. The shaded area is averaged to give the quasi-steady state current. } \label{fig:compHF}
\end{figure}

For interacting case, we calculate the real time spin current
through the ESR-QD by tVMPS method. The results were compared with
TD-HF results, see Fig.\ref{fig:compHF} for $U=1$. Although there
are discrepancies in the real time data, the TD-HF approximation
captures the overall behavior of the spin current. This is due to the
Kondo physics--which is missing from the TD-HF approach--does not
manifests itself here since the presence of large magnetic field
$B_1$ and Fermi surface splitting in the lead. Kondo effect may restored when $B_{1}$ is reduced
below Kondo temperature \cite{Hattori:2008p11172}. But small $B_{1}$ also reduce the chance of spin
flip process on the QD, thus reduce the spin current through it.
Surprisingly, the average of
physical quantities in the quasi-steady state of these two approaches are
in good agreement, see Fig.\ref{fig:varyVg}. Based on this
investigation, we validate that the HFA is a good approximation for
the qualitative investigation of spin-pump devises away from the
Kondo regime. However, caution must be taken when it is used to make
prediction on the real time dynamics. For example for the
$V_{g}=-1.0$ case, the transient state spin current of TD-HF is of
the opposite sign to the tVMPS prediction, thus it is quantitative
wrong.


The magnitude of spin current with the gate voltage is still a
two-peaked curve similar to the noninteracting case. Interaction has
different effects on the high and low filling regimes of the dot,
Fig\ref{fig:varyVg} (b). It does not modify the dot occupation and
the spin current dramatically for $V_{g}>0$. The peak of the spin
current remains at $V_{g}=1$ and the maximum value is only
suppressed by one percent up to $U=3$. However, since the average
dot occupation is larger than $1$ for $V_{g}<0$, the correlation
effect shifts the spin current peak downwards by $U$. This is a
manifestation of the Coulomb blockade effect.


The fluctuation of the density $\Delta = \langle\delta n_{\ua}\delta
n_{\da}\rangle =\langle n_{\ua}n_{\da}\rangle-\langle n_{\ua}\rangle
\langle n_{\da}\rangle$ is also calculated. It is a measure of the
accuracy of the HF approach. This quantity is conserved under the
particle-hole transformation. Thus two different gate voltages
satisfying $V_{g}+V_{g}^{\prime}=-U$ give the same value of
$\Delta$. This is respected in the tVMPS calculation. But the
discrepancy from the tVMPS is unrelated to it, {\it i.e.}, although
$V_{g}=1$ and $V_{g}=0$ has the same $\Delta$, the discrepancy of
the spin current from the tVMPS result is not the same. We find the
HF approximation is surprisingly good for the quasi-steady state
physical quantities even if the fluctuation is relatively large. The
fluctuation is always negative due to the Cauchy-Schwarz inequality,
indicating that the HFA overestimates the potential energy. The
absolute value of the fluctuation reaches its maximum $~0.2$ for
$-1< V_{g} < 0$. However, even for these cases, the HF data still
shows good agreement with the tVMPS data.


\section{Conclusion}
To conclude, we perform essentially exact real time calculation of
the spin current through an interacting ESR quantum dot. We
benchmark the essentially exact tDMRG result against those from
various analytical and approximate methods. From the extracted
average spin current, we obtain the Coulomb block shift of the spin
current peak, confirmed by the time-dependent Hartree-Fock
calculations. We find that the spin current attains its maximum for
a spin neutral quantum dot. The spin neutral condition is fulfilled
for two gate voltages where $V_{g}$ matches the magnitude of the
rotational magnetic field. The two spin current peaks respond
differently to the electronic correlation, the lower filling peak
shifts downwards by $U$ while the higher filling peak is nearly
unaffected. Comparison to the NEGF approach with an infinite lead
shows that the finite size effect does not affect the qualitative
behavior of the quasi-steady-state quantities. These results are
also compared with those of the TD-HF approach. It is shown that the
TD-HF gives accurate quasi-steady-state dot occupation and spin
current. However its prediction on the real time dynamics is
problematic.

\section{Acknowledgment}

The work is supported by NSFC and MOST of China. XCX is supported by
US-DOE under grant No. DE-FG02-04ER46124 and the C-SPIN center in
Oklahoma. LW and HJ thank P. Zhang for helpful discussions.

\bibliography{/Users/leiwang/Documents/Papers/papers}

\end{document}